\documentclass[apj]{emulateapj}
\usepackage{amsmath}
\usepackage{graphicx}
\usepackage[usenames,dvipsnames]{color}

\graphicspath{{img/}}
\newcommand{\MWO}{{\rm MWO}}
\newcommand{\SSS}{{\rm SSS}}
\newcommand{\SMARTS}{{\rm SMARTS}}
\newcommand{\HARPS}{{\rm HARPS}}
\newcommand{\CPS}{{\rm CPS}}
\newcommand{\APT}{{\rm APT}}
\newcommand{\Plong}{{\rm long}}
\newcommand{\Pshort}{{\rm short}}
\newcommand{\rot}{{\rm rot}}
\newcommand{\cyc}{{\rm cyc}}

\newcommand{\degree}{\ensuremath{^\circ}}
\newcommand{\anglemean}[1]{\ensuremath{\langle #1 \rangle}}
\newcommand{\revonecolor}{} 
\newcommand{\revone}[1]{{\revonecolor #1}} 

\begin{document}
\title{Sun-like magnetic cycles in the rapidly-rotating young solar analog HD 30495}
\author{Ricky Egeland\altaffilmark{1,2}, Travis S. Metcalfe\altaffilmark{3,4}, Jeffrey C. Hall\altaffilmark{5}, and Gregory W. Henry\altaffilmark{6}}

\altaffiltext{1}{High Altitude Observatory, National Center for Atmospheric Research‡, PO Box 3000, Boulder, CO 80307-3000, USA; \email{\mbox{egeland@ucar.edu}}}
\altaffiltext{2}{Department of Physics, Montana State University, Bozeman, MT 59717-3840, USA}
\altaffiltext{3}{Space Science Institute, 4750 Walnut St. Suite 205, Boulder CO 80301 USA}
\altaffiltext{4}{Stellar Astrophysics Centre, Department of Physics and Astronomy, Aarhus University, Ny Munkegade 120, DK-8000 Aarhus C, Denmark}
\altaffiltext{5}{Lowell Observatory, 1400 West Mars Hill Road, Flagstaff, AZ 86001, USA}
\altaffiltext{6}{Center of Excellence in Information Systems, Tennessee State University, 3500 John A. Merritt Blvd., Box 9501, Nashville, TN 37209, USA}

\begin{abstract}
A growing body of evidence suggests that multiple dynamo mechanisms
can drive magnetic variability on different timescales, not only in
the Sun but also in other stars. Many solar activity proxies exhibit a
quasi-biennial ($\sim$2 year) variation, which is superimposed upon
the dominant 11 year cycle. A well-characterized stellar sample
suggests at least two different relationships between rotation period
and cycle period, with some stars exhibiting long and short cycles
simultaneously. Within this sample, the solar cycle periods are
typical of a more rapidly rotating star, implying that the Sun might
be in a transitional state or that it has an unusual evolutionary
history. In this work, we present new and archival observations of
dual magnetic cycles in the young solar analog HD 30495, an $\sim$1
Gyr-old G1.5V star with a rotation period near 11 days. This star
falls squarely on the relationships established by the broader stellar
sample, with short-period variations at $\sim$1.7 years and a long
cycle of $\sim$12 years.  We measure three individual long-period
cycles and find durations ranging from 9.6--15.5 years.  \revone{We
find the short-term variability to be intermittent, but present
throughout the majority of the time series, though its occurrence
and amplitude are uncorrelated with the longer cycle.}  These
essentially solar-like variations occur in a Sun-like star with more
rapid rotation, though surface differential rotation measurements
leave open the possibility of a solar equivalence.
\end{abstract}


\section{Background and Characterization}
\label{sec:background}

Stellar magnetic activity cycles have been known in the Sun since
Schwabe and were shown to exist in other stars by \cite{Wilson:1978}
using an activity index derived from flux measurements in the Ca II HK
line cores performed at the Mount Wilson Observatory.  The Mount
Wilson survey from 1966--2003 remains the largest and longest campaign
investigating stellar activity, culminating with \cite{Baliunas:1995}
reporting on the variability of 111 stars, 52 of which were found to
demonstrate periodic behavior in a manner similar to our Sun.  This
periodic behavior is believed to be caused by a magnetic dynamo driven
by rotational motions of plasma in the stellar interior.  Using the
Mount Wilson data along with measurements of surface rotation,
\cite{Saar:1992} and \cite{Soon:1993} first reported two distinct
branches of cycling stars, the ``active'' and ``inactive'' branches
distinguished by their mean activity level and number of rotations per
cycle, and furthermore found some stars with multiple prominent
periodicities fit on both branches.  Observing \revone{a subset of high-quality cycle detections from the stellar
sample} of \cite{Saar:1999}, \cite{BohmVitense:2007} hypothesized that
these classes are due to separate dynamo mechanisms identifiable by
their interior shear layer, with the slow-rotating ``inactive'' branch
operating near the base of the convective envelope, and the
fast-rotating ``active'' branch working in rotational shear layers
closer to the surface.  Curiously, the Sun and its 11 yr cycle appear
to be a unique outlier in this small stellar sample, falling between
the two activity branches.  \cite{BohmVitense:2007} suggested that the
Sun could be in transition from one dominant dynamo mechanism to
another.

In addition to the 11 yr solar cycle, short-term quasi-periodic
variability has been observed in a number of solar phenomena.  Various
manifestations of the so-called quasi-biennial oscillation (QBO) of
0.6-4 yrs are reviewed in \cite{Bazilevskaya:2014} and
\cite{McIntosh:2015}.  QBOs are found in records of sunspot number and
area, magnetic field measurements, solar irradiance, and in
magnetically-sensitive phenomena such as filaments in H-alpha and
variations of field-sensitive lines such as Ca II and Mn I.  The QBO
also appears in eruptive phenomena -- flares, coronal mass ejections,
and solar energetic particle events -- which arise from magnetically
active regions.  Short period variations have also been observed in
the stars: \cite{Baliunas:1995} reported nine stars with significant
``secondary cycles''.  Six of these (and one new addition) were part
of the high-quality activity cycles of the \cite{Saar:1999} sample,
with the secondary cycle falling on the ``inactive'' branch.
\cite{Olah:2009} performed a time-frequency analysis of multi-decadal
photometry and Ca II emission for 20 stars and found 15 of them to
exhibit multiple cycles.  High-cadence SMARTS HK observations have
found short-period variations (1.6 yr) on $\iota$ Horologi
\citep{Metcalfe:2010} and $\epsilon$ Eridani (2.95 yr), another
dual-cycle star with a long-term cycle of 12.7 yr
\citep{Metcalfe:2013}.  \revone{\cite{Fares:2009} used Zeeman Doppler
  Imaging to observe a polarity-flipping cycle of $\sim$2 years
  in the fast-rotating F6 star $\tau$ Bootis, which had weak
  indications of a long-period activity cycle of 11.6 yr in
  \cite{Baliunas:1995}.  This short-period cycle is distinct from the
  solar QBO phenomena, which does not reverse magnetic polarity.  }

The origin of the solar QBO and its relation to the solar cycle is not
understood.  The discovery of periodic variations of 1.3 yrs in the
differential rotation of the deep interior revealed by helioseismology
\citep{Howe:2000} suggests that the QBO is sub-surface in origin and
may be an additional feature of a deep-interior dynamo process
responsible for the 11-year cycle \citep[see][and references
  therein]{Bazilevskaya:2014}.  \cite{McIntosh:2015} also points to a
deep interior process, inferring that this short-period variability is
driven by the interaction of two oppositely-signed magnetic activity
bands deep in the interior of each hemisphere.  Another possibility is the
distinct-dynamo scenario described above as an explanation for the two
activity branches in the \cite{Saar:1999}/\cite{BohmVitense:2007}
stellar sample. \revone{\cite{Fletcher:2010} find a $\sim$2 yr
  variation in the frequency shift of solar p-mode oscillations, and
  locate the origin of the variations to be \emph{below} the source of
  the 11-yr signal in the data.  They hypothesize that spatially
  distinct dynamo processes may be responsible for this phenomenon.}
\cite{Chowdhury:2009} suggested a non-dynamo origin
for QBOs: an instability caused by Rossby waves interacting with the
tachocline.

While stellar observations cannot match the level of detail in which
the solar QBO is observed, the varied physical conditions present in
other stars may have an impact on the manifestations of these
short-period oscillations that sheds additional light on their
origins.  In this work, we present new observations of variability in
HD 30495 (58 Eri), a nearby young solar analog that
demonstrates both a long-term activity cycle and a short-period
oscillation, which may be analogous to the solar cycle and
QBO.  The upper section of Table \ref{tab:prop} summarizes the
measured global properties of HD 30495. Photometry and spectroscopy
show the star to be essentially solar-like, leading previous authors
to study this star as a potential solar twin
\citep{CayrelDeStrobel:1996,PortoDeMello:2014}.  \cite{Gaidos:2000}
found rotational modulations in high-cadence Str{\"o}mgren $b$ and $y$
photometry to determine a rotation period of 11.3 days, roughly 2.3
times faster than the Sun.  Due to the process of magnetic braking,
older stars have slower rotations, giving the well-known
$P_\rot \approx t^{-0.5}$ age-rotation relationship
\citep{Skumanich:1972}.  The faster rotation of HD 30495 thereby
implies it is younger than the Sun, and by the age-rotation
relationship given in \cite{Barnes:2007} we obtain an age of about
$\sim$1 Gyr.  Observations of excess infrared flux attributed to a
diffuse and distant debris disk of $\sim$73 Earth-masses leftover from
formation \citep{Habing:2001} give further evidence for a young age.
Spectroscopic searches of similar ``Vega-like'' main-sequence objects
with excess infrared emission have ruled out the possibility of dense
concentrations of gas close to the star
\citep{Habing:2001,Liseau:1999}.  Based on these studies, in the
discussions that follow, we shall assume that the disk is not a
contributing factor to the observed magnetic signatures.

\cite{Baliunas:1995} previously searched for cyclic variability in HD
30495 using the Lomb-Scargle periodogram on the Mount Wilson $S$ time
series from 1966--1992, but classified it as ``Var'', defined as
``significant variability without pronounced periodicity'' and
$\anglemean{\sigma_S/S} \gtrsim 2\%$.  The mean activity level was
found to be high, $\anglemean{S} = 0.297$, as expected for fast
rotators (compare to the solar value from the same study
$\anglemean{S_\Sun} = 0.179$).  \cite{Hall:2007} $S$-index
measurements from the Solar Stellar Spectrograph (SSS) again found
stronger-than-solar activity levels, with $\anglemean{S} = 0.309$.
\cite{Hall:2009} used twelve years of precise Str{\"o}mgren $b$ and $y$
photometry from the Fairborn Automated Photometric Telescope (APT)
program \citep{Henry:1999}, finding a photometric variability for HD 30495 roughly six
times solar.  Furthermore, brightness was shown to decrease with
increased chromospheric activity, indicating that variability in the star's
brightness is dominated by dark spots, typical of fast rotators in that study.


Naively, and based solely on its similarity to the Sun and faster
rotation and, hence, presumably greater differential rotation, we might
expect HD 30495 to have a more vigorous dynamo, leading to higher
magnetic activity and a shorter activity cycle.  As we shall see, the
former is borne out by observations, while the latter is not.

\begin{deluxetable}{llr}
  \tabletypesize{\small}
  \tablecolumns{3}
  \tablewidth{0.45\textwidth}
  \tablecaption{HD 30495 Properties}
  \tablehead{\colhead{Property} & \colhead{Value} & \colhead{Reference}}
  \startdata
  Spectral type   & G1.5 V                      & (1) \\
  V               & $5.49 \pm 0.01$             & (2) \\
  B-V             & $0.632 \pm 0.006$           & (2) \\
  Parallax        & $75.32 \pm 0.36$ mas        & (2) \\
  $v \sin i$      & $4.1 \pm 0.8$ km s$^{-1}$   & (3) \\
  $T_{\rm eff}$   & $5826 \pm 48$ K             & (4) \\
  $\log g$        & $4.54 \pm 0.012$ dex        & (4) \\
  {[Fe/H]}        & $+0.005 \pm 0.029$ solar    & (4) \\
  Mass            & $1.02 \pm 0.01 \, M_\Sun$   & (4) \\
  Radius          & $0.898 \pm 0.013 \, R_\Sun$ & (4$^*$) \\
  Luminosity      & $0.837 \pm 0.037 \, L_\Sun$ & (4$^*$) \\
  Age             & $970 \pm 120$ Myr           & (5*) \\
  \tableline
  $\overline{P}_\rot$  & $11.36 \pm 0.17$ days  & \\
  $\Delta P$           & $0.59 \pm 0.05$ days   & \\
  $\Delta \Omega$      & $\gtrsim 1.67 \pm 0.15$ deg/day & \\
  $\sin i$             & $1.0 \pm 0.2$          & \\
  $i$                  & $\gtrsim 55.4^\circ$   & \\
  $P_{\cyc, \Plong}$   & $12.2 \pm [3.0]$ yr    & \\
  $A_{\cyc, \Plong}$   & $0.118 \pm [0.044]$    & \\
  $P_{\Pshort}$        & $1.67 \pm [0.35]$ yr   & \\
  $A_{\Pshort}$        & $0.066 \pm [0.028]$    & 
  \enddata
  \tablecomments{References: (1) \cite{Gray:2006} (2) Hipparcos,
    \cite{Perryman:1997, vanLeeuwen:2007} (3) \cite{Gaidos:2000} (4)
    \cite{Baumann:2010} (4*) Derived from \cite{Baumann:2010}
    measurements (5*) Derived from \cite{Barnes:2007} age-rotation
    relationship. The lower section of the table are measurements found
    in this work \revone{Quantities in brackets represent one half of
    the observed range of values.}
  }
  \label{tab:prop}
\end{deluxetable}

\section{Activity Analysis}
\label{sec:analysis}

\begin{figure*}
\centering
    \includegraphics[width=\textwidth]{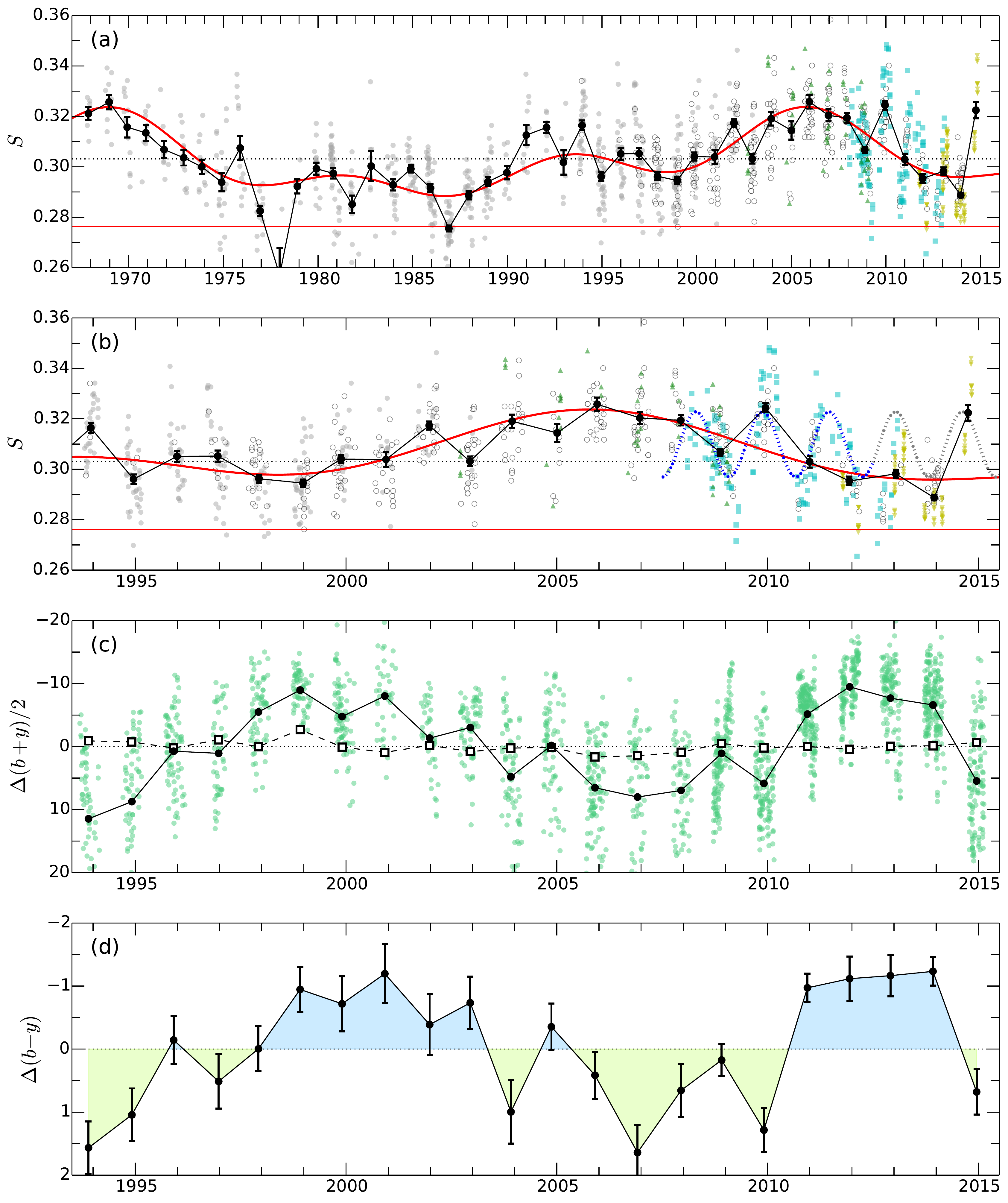}
    \caption{(a) HD 30495 combined $S$-index time series, including
      data from MWO ({\color{Gray}$\bullet$}), SSS ($\circ$), SMARTS
      ({\color{Cyan}$\blacksquare$}), CPS
      ({\color{ForestGreen}$\blacktriangle$}), and HARPS
      ({\color{Yellow}$\blacktriangledown$}), along with seasonal
      means ($\bullet$), with error bars representing the error of the
      mean.  The red curve is a 3-component sine wave model of the
      stellar cycles, \revone{while the horizontal red line is our reference
      point for global activity minimum.}  (b) Zoomed portion
      highlighting higher-cadence SMARTS data, with a $P  =
      1.58$ yr sine wave plotted for comparison (blue dashed
      curve). (c) APT differential photometry brightness measurements
      in the combined Str{\"o}mgren $b$ and $y$ bands, in
      milli-magnitudes. Differences shown are HD 30495 nightly
      measurements ({\color{YellowGreen}$\bullet$}) and seasonal means
      ($\bullet$) with respect to the comparison stars, as well as the
      difference between the two comparison stars ($\square$) (d)
      Seasonal mean differential brightness difference in the $b$ and
      $y$ bands, in milli-magnitudes, with colored regions indicating
      brighter $b$ (blue) and $y$ (green) emission.  A horizontal
      dotted line indicates the grand mean in all panels.
      \revone{Magnitude scales are inverted such that brightness
        increases in the upward direction.  The observations shown in
        this figure are available in the electronic version of this
        publication.  See tables \ref{tab:obs_S} and \ref{tab:obs_by}
        in the Appendix.}}
    \label{fig:timeseries}
\end{figure*}

We analyze a combined 47-year time-series of the Mount Wilson
$S$-index shown in Fig \ref{fig:timeseries}.  This dimensionless index
is defined as the ratio of the core emission in the Fraunhofer H and K
lines of Ca II with the nearby continuum regions, as measured by the
HKP-1 and later HKP-2 photometers at Mount Wilson Observatory (MWO)
\citep{Wilson:1968, Vaughan:1978}.  Ca II H \& K global-scale emission
reversals are a signature of departure from radiative equilibrium, a
defining feature of a chromosphere, and must be due to magnetic
non-thermal heating mechanisms.  Due to the unsurpassed duration and
breadth of the Mount Wilson survey, $S$, a measure on an
\emph{instrumental scale}, has become the \emph{de facto} standard
for measuring stellar magnetic activity, and subsequent surveys
have calibrated to the Mount Wilson scale.  $S$ \revone{is dependent on
  stellar properties such as temperature, surface gravity, and
  composition}, which precludes its use for directly comparing
activity levels of a \revone{heterogeneous ensemble}.  As we focus our
analysis on a single star, conversion to a corrected quantity (e.g.,
$R'_{HK}$; \cite{Noyes:1984}) is not necessary.

\begin{figure*}[ht]
    \includegraphics[width=\textwidth]{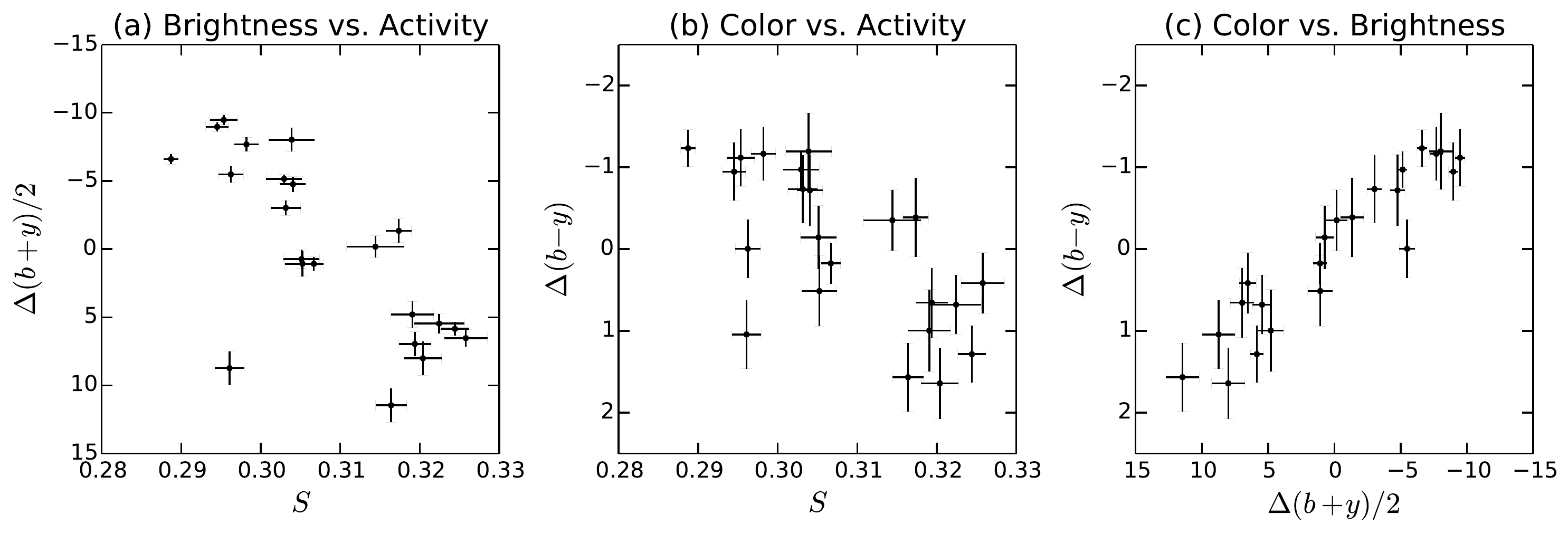}
    \caption{Correlations among seasonal means of activity, brightness
      and color from the time series of Figure \ref{fig:timeseries}.
      Error bars indicate the error in the mean.  Magnitude scales are
      in milli-magnitudes and are inverted such that brightness
      increases in the upward/rightward directions.}
    \label{fig:correlations}
\end{figure*}

The combined $S$ time series in Figure \ref{fig:timeseries}(a)
contains 1285 measurements from five different instruments.  The
majority of the measurements (624 measurements from 1967-2003) come
from the original Mount Wilson survey, calibrated as
described in \cite{Baliunas:1995}.  \revone{We assumed a uniform measurement error of
3\% of the mean for this time series}, near the upper limit quoted by
\cite{Wilson:1968}.  The next largest portion of the measurements are
from the Solar-Stellar Spectrograph (SSS) at Lowell Observatory (364
measurements from 1993-2014) \citep{Hall:1995}, taking the time series
from the beginning of the Mount Wilson survey to present day.
\revone{SSS obtains $R \approx 10,000$ at H$\alpha$ spectra, and $S$
  is derived by apeproximating the Ca II H, K and continuum bandpasses
  used by the MWO instrument.  These data are then calibrated to the
  Mount Wilson instrumental scale using long-duration means of common
  targets.}  A typical measurement error of 2.4\% was
estimated using photon statistics in the K line core and detector
properties.  Observations from the SMARTS Southern HK survey using the
\emph{RC Spec} $R \approx 2500$ spectrograph at 1.5-m telescope at
CTIO are the third largest contribution (140 measurements from
2008--2013), and though of shorter duration, this queue-scheduled time series is not
plagued by the large seasonal gaps of the other surveys, allowing
short-period variation to be better determined.  These data were
calibrated to the Mount Wilson scale via common observations with SSS
targets, as described in \cite{Metcalfe:2010}.  An additional 108
measurements from 2011--2015 derived from HARPS \revone{$R
\approx 120,000$ spectra from a solar twin planet search
\citep{Ramirez:2014,Bedell:2015}, again calibrated to the MWO scale
using common targets, as described in \citep{Lovis:2011}.}  Finally, we
add 49 observations from 2002--2008 derived from $R \approx 55,000$
spectra of the Hamilton Spectrometer at Lick Observatory.  These
observations\footnote{http://cdsarc.u-strasbg.fr/viz-bin/Cat?J/ApJ/725/875}
are part of the California Planet Search (CPS) and were similarly
calibrated to the MWO scale using common targets
\citep{Isaacson:2010}.

Though each of these time series used a global calibration to the
Mount Wilson scale using long-term means of commonly observed targets,
visual inspection of the combined time series revealed obvious
discontinuities and differences in scale.  This is likely due to the
fact that the \emph{global} calibration involves a compromise linear
fit among all targets, while scatter about that fit reveals error in
the calibration that would result in a discontinuity in any individual
target.  We applied a simple calibration that assumes
overlapping periods of two different time series ought to agree on the
mean for that period.  To calibrate $S$ to the scale of $S_0$, the
mean value over the period of overlap, $\overline{S}$ and
$\overline{S}_0$ were calculated, and a scaling factor $C =
\overline{S}_0 / \overline{S}$ was derived.  The resulting calibrated
time series $S' = C S$ then has an equivalent mean value over the
overlapping period to the base series $S_0$.  The resulting scaling
factors were $C(\SSS\rightarrow\MWO) = 1.015$,
$C(\SMARTS\rightarrow\SSS') = 1.067$, $C(\HARPS\rightarrow\SSS') =
1.074$, $C(\CPS\rightarrow\SSS') = 1.098$.  The SMARTS, HARPS, and CPS
time series were scaled using overlapping portions of the
post-calibration SSS time series, therefore their overall scaling is
multiplied by $C(\SSS\rightarrow\MWO)$.  This calibration removed
obvious discontinuities in the combined time series and reduced the
standard deviation by 3.8\%.  The final combined time series has a
grand mean $\overline{S} = 0.303$ and a standard deviation $\sigma =
0.0167$.  Seasonal means for the combined time series are shown as black
circles in Figure \ref{fig:timeseries}(a).  Following these seasonal
means, clear cyclic behavior is visible, emphasized by the cycle model
(red curve) described below.


We also examined the 22-yr time series of differential photometry
acquired with the T4 0.75 meter Automatic Photoelectric Telescope (APT)
at Fairborn Observatory \citep{Henry:1999}, shown in Figure \ref{fig:timeseries}(c).  These
measurements, made in the Str{\"o}mgren $b$ (467 nm) and $y$ (547 nm)
bands, are a difference with respect to the mean brightness of two
stable comparison stars, HD 31414 and HD 30606. The differential
measurements in the $b$ and $y$ bands are then averaged to $(b + y)/2$
to create a ``$by$'' band that increases the signal to noise
ratio. The unimportant mean difference is subtracted from the time
series.  The stability of the comparison stars is demonstrated in the
seasonal mean of their brightness difference in the $by$ band, shown
as white squares in Figure \ref{fig:timeseries}(c), with a standard
deviation $\sigma = 0.00093$ mag.  HD 30495 $by$ brightness is
strongly variable ($\sigma = 0.0065$ mag) and out-of-phase with the
$S$-index shown in Figure \ref{fig:timeseries}(b). \revone{A rank correlation
test between $S$ and $by$ seasonal means shows 99.98\% significance in
the correlation, which is plotted in Figure \ref{fig:correlations}(a).}  This is interpreted as evidence the star's
brightness variations are dominated by dark spots, which are more
prevalent during times of activity maximum.  Figure
\ref{fig:timeseries}(d) plots the $\Delta (b - y)$ color index where
blue shading indicates negative color index and green shading
indicates positive color index.  \revone{Comparing panels (b), (c) and
  (d) of Figure \ref{fig:timeseries}, we see that HD 30495 gets bluer
  as it gets brighter (activity minimum) and redder as it gets fainter
  (activity maximum).  This is shown again more clearly in Figure
  \ref{fig:correlations} (b) and (c), in particular the remarkably tight
  color-brightness correlation.  We interpret this color shift as an
  increase in surface temperature during times of activity minima,
  due to the reduction of cool spots on the surface.}

\begin{figure*}
    \includegraphics[width=\textwidth]{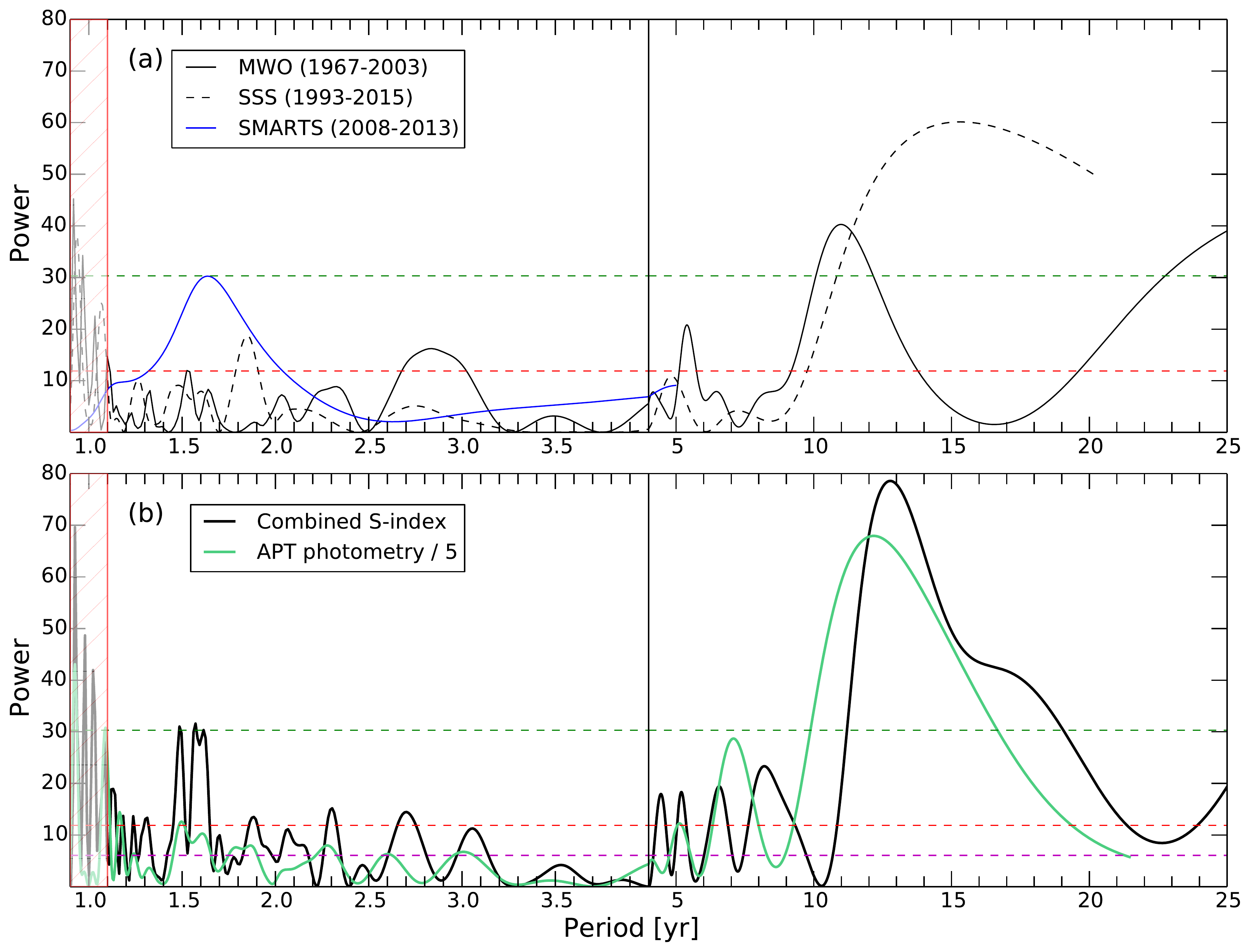}
    \caption{Lomb-Scargle periodograms from the time series of
      Figure \ref{fig:timeseries}.  Panel (a) contains the result from
      single-instrument S-index surveys and panel (b) the combined
      S-index time series, as well as the APT photometry of Figure
      \ref{fig:timeseries}(c).  Note the division in the period scale.
      The hatched region near $P \approx 1$ yr contains artifacts of
      the seasonal sampling.  The green and red horizontal dashed
      lines are the ``excellent'' and ``poor'' significance thresholds
      for the S-index periodograms, as defined in
      \cite{Baliunas:1995}.  Note that the APT periodogram is scaled
      down by a factor of five for easy comparison and the magenta
      horizontal line is the ``excellent'' threshold for that series.}
    \label{fig:pgrams}
\end{figure*}


We computed the Lomb-Scargle periodogram (\cite{Scargle:1982},
\cite{Horne:1986}) from our time series to find statistically
significant periodicities in the data, with the results shown in
Figure \ref{fig:pgrams}. To verify the robustness of the peaks, we
compare the periodogram of the combined time series (thick black line)
to those of the individual MWO, SSS, and SMARTS series over shorter
intervals, as well as to the periodogram of the $(b+y)/2$ photometry.
Not shown in the figure are the large peaks beyond 25 years in the MWO
and combined $S$-index periodograms, which are most likely due to the
windowing of the entire time series, not true physical variation.  The
hatched regions of $P < 1.1$ on the left side of periodograms contain
a number of large peaks near 1 year, which are aliases due to the
seasonal sampling in our time series.  We verified these are all
aliases by obtaining a least-squares fit of the data to a sine wave
with a period set by one of the $\sim$1 yr peaks, then subtracting
that signal from the time series and re-computing the periodogram.
The new periodogram would no longer contain the $\sim$1 yr peak, and a
corresponding low-frequency peak would be removed as well.  This
established a symmetry between the low-frequency peaks and these
$\sim$1 year peaks, and as a result we do not consider any peak $<$
1.1 years to be physical.  \revone{(See also Figure
\ref{fig:cyclepgrams}(c), in which spurious $\sim$1 yr peaks are
found in the periodogram of a signal of pure sine waves of lower
frequency.)}


Following \cite{Horne:1986}, we calculate the
``False Alarm Probability'' (FAP) threshold:

\begin{equation}
  \label{eq:fap}
   z = - \ln \left(1 - \left(1 - F \right)^{1/N_i} \right)
\end{equation}

where $F$ is the probability that there exists a peak of height $z$ at
any frequency due to random Gaussian noise in the signal, and $N_i$ is
the number of independent frequencies in the time series.  We computed
$N_i$ by generating 5000 random time series with the same sampling
times of our data, generating a probability distribution for the
maximum peak $z$, and fitting this distribution to equation
\eqref{eq:fap} inverted for $F$, with $N_i$ as the free parameter.
The upper threshold (green line) shown in Figure \ref{fig:pgrams}
corresponds to $F = 10^{-11}$, the threshold for an ``excellent''
cycle in \cite{Baliunas:1995}, and the lower threshold (red line) is
for $F = 10^{-3}$ (99.9\% significance), the minimum requirement for a
``poor'' cycle in that work.  The FAP thresholds shown are those
computed for the combined time series, however they are
similar to those obtained for the individual component time series,
being only slightly more stringent.

The uncertainties in peak positions were estimated using a Monte Carlo
method.  In each trial, each time series measurement is randomly
sampled from a Gaussian distribution defined by that measurement's
value and uncertainty.  Then, a periodogram is computed and the new
peak position saved.  By running 5000 trials, an approximately
Gaussian distribution of peak positions is obtained, and the
uncertainty is estimated as its standard deviation.


In the combined time series we found four resolved peaks above the
``excellent'' threshold: a long period peak $P_\Plong = 12.77 \pm
0.09$ yr, and a cluster of three short-period peaks at $P_{\Pshort,1}
= 1.572$, $P_{\Pshort,2} = 1.486$, and $P_{\Pshort,3} = 1.615$ yr
($\sigma_\Pshort \approx 0.003$).  The $P_\Plong$ peak is found
between nearby significant peaks found in the MWO ($P_{\Plong, \MWO} =
10.7$ yr) and SSS ($P_{\Plong, \SSS} = 15.3$ yr) time series, and
nearby the $P_{\Plong, \APT} = 12.2$ peak from $by$ photometry.  The
spread in periods from the earlier MWO data to the later SSS data
indicates that, like the Sun, the long-term cycle is only
quasi-periodic, and the duration of each individual cycle varies.  The
$\sim$13 yr peak has a protruding ``shoulder'' on its right side, which
is due to an unresolved peak near 17 yr.  This peak was resolved in
$\sim$25\% of the Monte Carlo trials done to determine the
uncertainty in peak positions, allowing us to measure a mean value of
16.9 yr.  This $\sim$17 year period \revone{and another large peak at
  $\sim$37 yr were found to be artifacts of the amplitude structure
  and/or the duration of the time series (data window).  We verified
  this by computing periodograms of various fractions of the data
  window (e.g. 2/3 to 1/2 of the total duration) at various offsets and
  noting that the $\sim$17 yr peak disappears in all cases and the
  $\sim$37 yr peak shifts close to the duration of the new window.}


Figure \ref{fig:pgrams}(b) shows the
short-period peaks are almost perfectly matched by two peaks (1.49 yr
and 1.61 yr) in the APT periodogram.  \revone{We also find a broad corresponding
``excellent''-class peak at 1.63 yr in the SMARTS time series and again in
the HARPS series at 1.75 yr.  A less significant ``poor''-class peak
at 1.85 yr is in the SSS data and 1.53 yr in the MWO data.}  The spread
in values indicates that these short-period variations are not of a
constant frequency\revone{, which we investigate in detail in Section
\ref{sec:time-frequency}}.


We find that both the long and short-period signals are found
consistently in several distinct $S$-index time series of different
time intervals, as well as in the APT differential photometry, a
measurement using very different observation methods to sample a
physically distinct region of the star.  This is strong evidence for
the co-existence of variability on different time scales and in
distinct regions of the stellar atmosphere, analogous to the solar
observations of the 11-year cycle and quasi-biennial oscillation.

\section{Simple Cycle Model}
\label{sec:cycle}

In the Sun, each occurrence of rising and falling activity is numbered
and, somewhat confusingly, referred to as a ``cycle'', with the
current episode denoted as ``cycle 24''.  Properties of each cycle
such as duration, amplitude, and shape are measured and found to vary.
We wish to similarly decompose the $\sim$12 yr periodic signal of
HD 30495 into individual cycles and measure their properties.  For
the Sun, this decomposition is typically done by identifying cycle
minima in a smoothed time series of a proxy such as sunspot number,
e.g. the 13-month boxcar smoothing of the monthly averages, and then
using the minima as delimiters for each cycle \citep{Hathaway:2010}.  The
seasonal gaps in stellar time series do not allow us to use this same
prescription.  Instead, we construct an idealized smoothed model of
the time series as a superposition of \revone{low-frequency} sine
waves:

\begin{equation}
  \label{eq:sine}
  S(t)_i = A_i \sin \left( \frac{2 \pi}{ P_i } \left( t + \phi_i \right) \right) + y_i
\end{equation}

where $P_i$ is a low-frequency period \revone{from a periodogram
  analysis}, and the amplitude $A_i$, phase $\phi_i$ and offset $y_i$
are found using a least squares optimization of this model to the
mean-subtracted data.  The final model is simply:

\begin{equation}
  \label{eq:sumsine}
  S(t) = \sum_i^{N_P} S(t)_i + \overline{S}
\end{equation}

where $N_P$ is the number of component sine waves and $\overline{S}$
is the original mean $S$-index.  

\begin{figure}[ht]
    \includegraphics[width=0.48\textwidth]{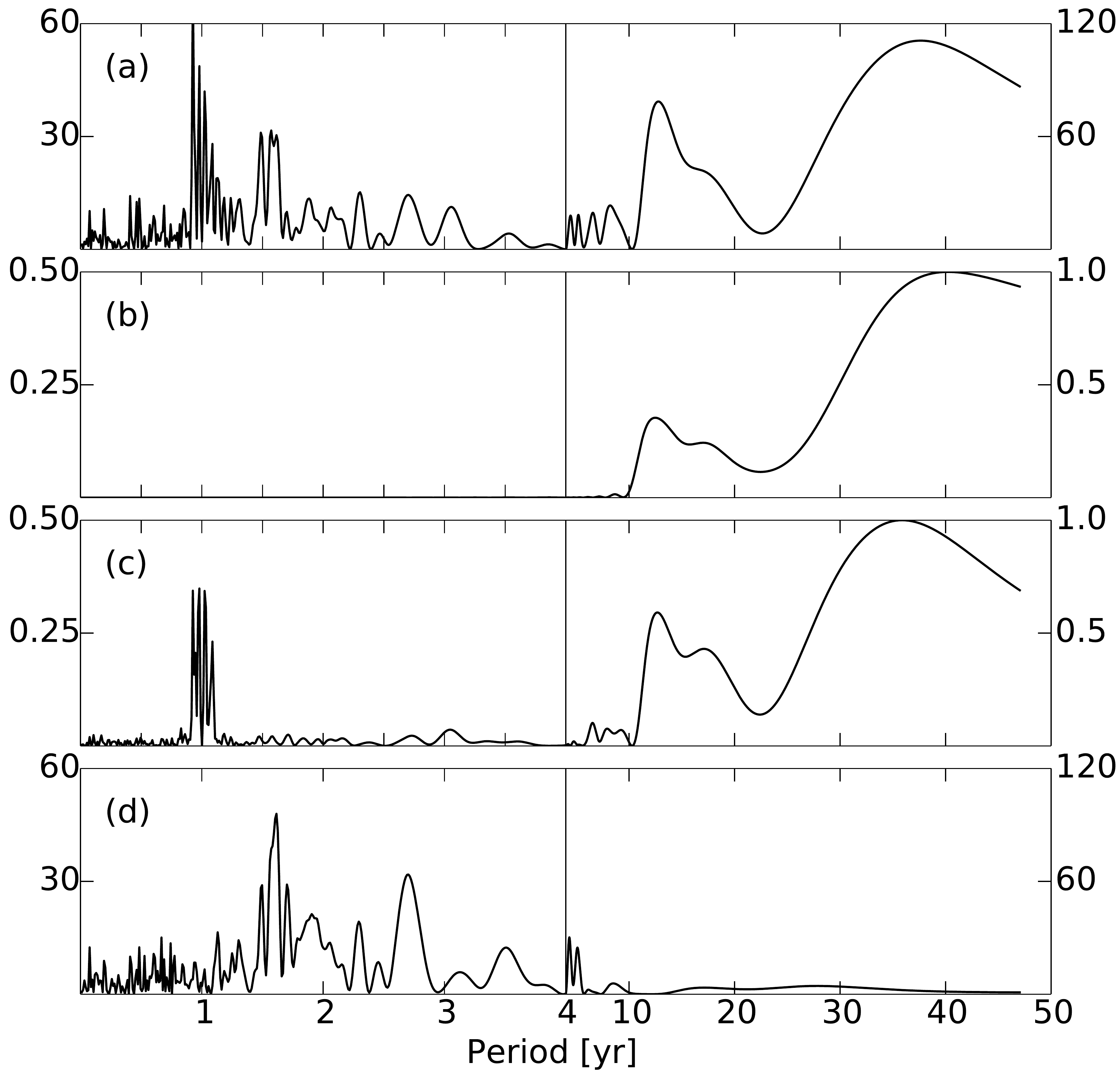}
    \caption{Lomb-Scargle periodogram of(a) the original composite
      $S$-index time series (b)
      3-component cycle model with equal-spaced sampling (c)
      3-component cycle model with the same sampling as the original
      data (d) the residual of the original data minus the cycle
      model.  Note that period and power scales change at the 4-yr
      mark; the left and right y-axis give the power scale for that
      side.  Periodograms of the cycle model are normalized to 1.
      }
    \label{fig:cyclepgrams}
\end{figure}

{\revonecolor
We obtain the parameters of equation \eqref{eq:sine} by iteratively
finding the lowest-frequency period in a Lomb-Scargle periodogram of
the composite $S$-index time series, fitting the sine to the data
using a least-squares optimization, and subtracting the result from the
data before computing the next periodogram.  We carry out three such
iterations in order to find the three low-frequency components
revealed by the periodogram of the original dataset.  The resulting
model is akin to the result of a low-pass filter on the data.
Indeed, subtracting the model from the data and computing the periodogram
reveals that we have effectively removed most of the
low-frequency power, as can be seen in Figure \ref{fig:cyclepgrams}.
The parameter set obtained from the fits of
equation \eqref{eq:sine} were $P = \{37.6, 18.8, 12.2 \}$, $A =
\{11.4, 4.68, 6.64 \} \times 10^{-3}$, $\phi = \{-0.103, 5.93,
5.45\}$, and $y = \{-2.47, -0.0519, 1.26 \} \times 10^{-3}$.
}

\begin{deluxetable}{cccccc}[ht]
  \tabletypesize{\small}
  \tablecolumns{6}
  \tablewidth{0.45\textwidth}
  \tablecaption{HD 30495 Cycle Properties}
  \tablehead{\colhead{Cycle} & \colhead{Start} & \colhead{Max} & \colhead{Duration} & \colhead{$S_{\rm max}$} & \colhead{$A_{\rm cyc}$}}
  \startdata
    0 & (1961.7) & 1969.0   & (15.3) & 0.324 & 0.156 \\  
    1 & 1977.1   & 1981.2   &  9.6   & 0.297 & 0.067 \\  
    2 & 1986.7   & 1993.7   & 11.7   & 0.305 & 0.095 \\
    3 & 1998.4   & 2005.8   & 15.5   & 0.324 & 0.156 \\
    4 & 2013.9   & --       & --     & --    & --
  \enddata
\label{tab:cycles}
\end{deluxetable}

Using this model we characterized the individual cycles of HD 30495 as
shown in Table \ref{tab:cycles}.  Quantities in parenthesis are based
on an extrapolation beyond the data, and should be treated with
caution.  We see a spread in cycle durations from 9.6 to 15.5 yr, with
a mean value of the fully observed cycles 1--3 of 12.2 yr.  This value
is close to the $P_\Plong$ = 12.77 yr peak in the periodogram of
Figure \ref{fig:pgrams}(b).  \revone{We shall adopt the mean value of 12.2 yr as our
best estimate for the mean cycle period for HD 30495.}  The spread of individual
cycle durations from the model gives us an estimate of the variability
of cycle durations, $\Delta P_\Plong /2$ = 3.0 yr.  The increase in
cycle duration from cycle 1 to 3 obtained from our model agrees with
the trend observed in the periodograms, with the earlier MWO series
which includes cycles 1 and 2 having $P_{\Plong, \MWO} = 10.7$, and
the later SSS series which only includes cycle 3 having $P_{\Plong,
  \SSS} = 15.3$ (see Figure \ref{fig:pgrams}(a)).  \revone{The cycle
  model components at $P$ = 37.6 and 18.8 yr are artifacts of the
  amplitude structure in the time series and the data window, as
  discussed above.  These components are necessary, however, to
  reconstruct the amplitude and cycle duration variability present in
  the original time series.  For example, the amplitude $A = 6.64
  \times 10^{-3}$ from the $P = 12.2$ yr least-squares fit to the
  original data is not representative of the amplitude of the
  individual cycles seen in Figure \ref{fig:timeseries}(a), being at
  least a factor 2 too low.  Only when the other two low-frequency
  components are included does the model amplitude more closely match
  the data.}

{ \revonecolor
  The relative cycle amplitudes from the model are shown in column 6
  of Table \ref{tab:cycles}, defined as $A_{\rm cyc} = \Delta S /
  \overline{S}$ following \cite{Soon:1994} and \cite{Saar:2002} studies
  of cycle amplitudes from the Mount Wilson program.  Determining
  $\Delta S = S_{\rm max} - S_{\rm ref}$ requires us to select a
  reference point that approximates the lowest possible activity level
  for this star.  Our cycle model does not effectively model the
  depth of the minima, as can be seen in Figure
  \ref{fig:timeseries}(a), which precludes its use for setting $S_{\rm
    ref}$.  Instead, we choose $S_{\rm ref}$ to be the lowest seasonal median
  with more than 3 measurements, $S_{\rm ref} = 0.2763$ from the
  1986-1987 season.  This definition of activity minimum avoids too
  much sensitivity to outliers, while being low enough that
  only 3.6\% of the data lie below this point.  With our choice of
  $S_{\rm ref}$ and $\overline{S} = 0.303$
  we find $A_{\rm cyc}$ ranging from 0.067 to 0.156, and a
  mean amplitude $\overline{A_{\rm cyc}} = 0.118$.  We find an
  increase in cycle amplitude from cycles 1 to 3, which occurs in
  parallel with the increase in cycle duration.  Note that for the Sun
  the amplitude of a subsequent cycle is negatively correlated with
  cycle period \citep{Hathaway:2010}.  The two transitions of fully
  observed cycles for HD 30495 indicate a positive correlation, though
  obviously no firm conclusions can be drawn from so little data.
}

\begin{figure*}[ht]
    \includegraphics[width=\textwidth]{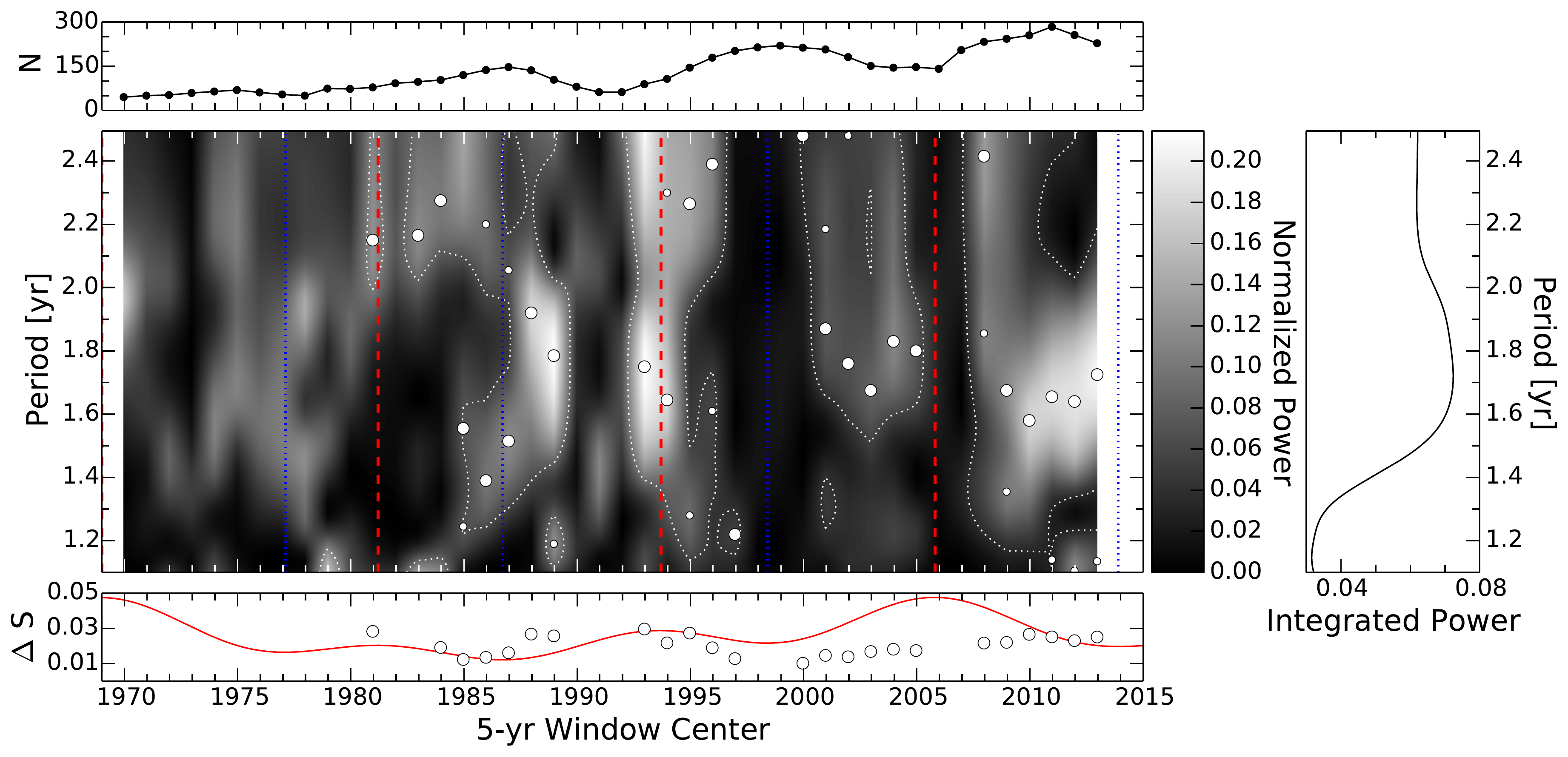}
    \caption{Short-Time Lomb-Scargle analysis.  A Lomb-Scargle
      periodogram is computed for 5-yr moving windows in 1 yr
      increments of the composite $S$-index time series for HD 30495.
      The contour plot gives the periodogram power normalized by the
      number of data points as a function of time and period, with the
      99.9\% significance contour highlighted as a dotted white line.
      The indicated time is for the center of the 5-yr window.  In
      each window, the highest peak above the 99.9\% threshold is
      found and plotted as a large open circle, while a secondary
      significant peak, if present, is plotted as a small open circle.
      Vertical blue and red dashed lines indicate the minima and
      maxima in the long-term cycle model.The top panel indicates the
      number of data points in each 5-yr window.  The right panel
      gives the integrated normalized power for all windows.  The
      bottom panel plots the amplitude $\Delta S$ of sine fits of the
      significant peak periods in the windowed data as open circles,
      with the amplitude of the 3-component cycle model in red.  }
    \label{fig:stls}
\end{figure*}

Robustness of the cycle model was examined by running Monte Carlo
simulations \revone{with the nightly measurements resampled from
  within their estimated uncertainties for each trial, then repeating
  the iterative procedure described above to compute the cycle model
  periods, amplitudes, phases, and offsets.}  The resulting minima
positions, maxima positions, and amplitudes were gathered and the
standard deviation of those distributions computed.  The times of
minima and maxima were found to be relatively robust forming a
Gaussian distribution with a standard deviation of $\sigma = 0.25$ yr.
\revone{The amplitudes were also fairly robust, forming a Gaussian
  distribution with $\sigma = 8.6 \times 10^{-4}$.  We use this to
  estimate the uncertainty in $\Delta S$ to be about 7\%, which
  dominates the uncertainty of our relative amplitudes $A_{\rm cyc} =
  \Delta S / \overline{S}$.}

{
\revonecolor

\section{Short-Period Time-Frequency Analysis}
\label{sec:time-frequency}

A sine fit with $P = 1.58$ was found for the combined time series for
a 5-year window centered at 2010.0 is shown in Figure
\ref{fig:timeseries}(b), with the curve colored cyan within the
fitting window and colored gray outside of the window.  Comparing this
sine wave to the data visually demonstrates the short-period
variation, especially visible during the epoch of SMARTS measurements
(cyan points).
Even in this short segment we notice that the data goes out-of-step
with the sine curve after 2012.  This shows by example the quasi-periodic
nature of the short-period variations, which was also evident in the
triplet of peaks near $P \approx 1.6$ yr in Figure \ref{fig:pgrams}(b).

To investigate possible coupling between the short- and long-period
variations found above, we performed a periodogram analysis on a 5-yr
moving window of our combined $S$-index time series, the results of
which are shown in Figure \ref{fig:stls}.  First, we removed the
low-frequency components from the data by subtracting the cycle model
described in Section \ref{sec:cycle}.  Comparing Figure
\ref{fig:cyclepgrams} panels (a) and (d), we see that this procedure
amplifies the high-frequency peaks in the residual periodogram.
However, we found the results of this analysis are qualitatively the
same when working with the original composite time series.  Next, the
residual time series is divided into 1-yr bins, which contain the
entirety of the MWO/SSS seasonal observations.  Every set of five
consecutive bins are then subjected to a Lomb-Scargle periodogram
analysis as described in Section \ref{sec:analysis}, searching for
periods from 1.1 to 2.5 years.  Five year windows were chosen as a
sufficient time period to capture multiple oscillations of the $P
\approx 1.6$ yr variability detected in Figure \ref{fig:pgrams}(b).
The period search cutoff at 1.1 yr is to avoid aliasing issues related
to the seasonal sampling, and the 2.5 yr cutoff is to avoid strong
signals associated with the duration of the window.  In total, 44
periodograms were calculated as well as a 0.001 (99.9\% significance)
false alarm probability threshold based on a Monte Carlo analysis as
described in Section \ref{sec:analysis}.  Up to two significant peaks
were extracted from these periodograms, and a least-squares fit of a
sine function to the data window is done to determine the amplitude of
that signal.  In Figure \ref{fig:stls} the normalized periodogram
power is plotted as a contour area plot in the time-period plane, and
the positions of significant peaks are plotted as open data points.
The 99.9\% significance contour is plotted as a dotted white line.

It is important to note that higher periodogram power is possible in
time series with a larger number of data points, so that periodogram
power does not scale equally from one window to the next.  To
check this behavior, we generated unevenly sampled time series of a
sine function by randomly removing 50\% of the data points and
calculated the Lomb-Scargle periodogram, finding that periodogram
power is proportional to the number of data points.  To correct for
this, we normalized the periodogram power by the number of
data points in the window.  Differences in periodogram power between
windows are then due to differences in amplitude of the underlying
signal.  The number of data points in each window is plotted in the
top panel of Figure \ref{fig:stls} for reference.  The 99.9\%
significance threshold is computed separately for each window, so the
contour position is correct despite the difference in the number of
data points.

Figure \ref{fig:stls} reveals that variability near $P_\Pshort \approx
1.6$ yr is intermittent with peak periods occuring across the full
range of periods analyzed.
Significant peak variability in the 1.4-1.8 yr range begins in 5-yr
windows centered at 1985, 1993, 2000, and 2009, with each episode
lasting for 3-5 years.  The 1985 and 2009 episodes precede the cycle
2 and 4 minima in the long-period cycle, respectively, while the 1993 and 2002 episodes roughly coincide
with cycle maxima.  In contrast to the minima of cycles 2 and 4, the
cycle 1 and 3 minima are devoid of short period variability.  Peaks
are often found from $\sim$2.2 to $\sim$2.4 yr as well, which is
confirmed in the plot of the power integrated over all windows.  Here
the broad peak at $\sim$1.7 yr is notably shifted from the cluster of
narrow peaks in the periodogram of Figure \ref{fig:pgrams}(b). This may be
due to varying phase in intermittent short- period signals leading to
interference effects in the full-time-series periodogram.  We will
take the mean of the detected peak periods $<$ 2.0 yr as our best
estimate of the short-period signal, $P_\Pshort = 1.67 \pm [0.35]$,
where the quantity in brackets is half the observed range of peak
periods.

We analyzed the peak-to-peak amplitudes $\Delta S$ of sine fits to the
data with the significant peak periods in the range of 1.1 to 2.0 yr
and found them to range from 0.012 to 0.030, with a mean of 0.020.
The average short-period relative amplitude $A_{\rm short} = \Delta S/
\overline{S}$ is then $0.066 \pm [0.028]$, roughly half of the average
long-period amplitude but nearly equal to relatively low amplitude of
cycle 1, as deduced from our cycle model.  We performed a
rank-correlation test between the short-period amplitudes and the
long-period cycle model, but no significant correlation was found.

From the above observations we conclude that there is no clear
association between the long-period cycle and the episodic
short-period variations.  The presence or absence of the short-period
variations are found in all phases of the long-term cycle, and the
amplitudes are not correlated with the long-term cycle amplitude.

} 

\section{Rotation}

We repeated the rotation measurements done for six seasons of APT
photometry in \cite{Gaidos:2000} using the current 22-season record.
The dense sampling of the APT program allows the detection of
rotational modulations due to the transit of spots on the stellar
photosphere.  The time series is broken into individual seasons
containing 55--185 measurements over the course of 150--200 days.  From
each season's time series a Lomb-Scargle periodogram was computed
looking for rotational periods from 2--25 days.  Peaks passing the
99.9\% significance level are taken as a signal of rotation.
Uncertainty in the peak position was calculated using the Monte Carlo
method described in Section \ref{sec:analysis} and ranged from 0.014
to 0.2 days.  In total, rotational periods were detected in 17 of 22
seasons.  Taking the mean, we find the average $P_\rot = 11.36 \pm
0.17$ d.  This result is within the uncertainty of the previous
\cite{Gaidos:2000} measurement, as well as that of
\cite{Baliunas:1996b} who obtained rotation from a densly-sampled
season of MWO $S$-index observations. The range of the precisely
determined rotations are from $10.970 \pm 0.028$ to $11.560 \pm 0.023$
days, giving $\Delta P = 0.59 \pm 0.05$ d.  This is reduced
from the \cite{Gaidos:2000} value of $\Delta P = 1.0$ d, due to the
fact that we could not reproduce a significant 10.5 d detection in
season 6. $\Delta P$ is interpreted as a sign of surface differential
rotation, due to transiting spots at various latitudes.  Our
measurements range over one and a half stellar cycles, which increases
the confidence that we have explored the full range of rotational
periods which can be sampled with this technique.  However, due to the
unknown latitude ranges of the spots on the star, the measurement
provides only a lower bound on the true equator-to-pole surface
differential rotation.

\begin{figure}[t]
    \includegraphics[width=0.45\textwidth]{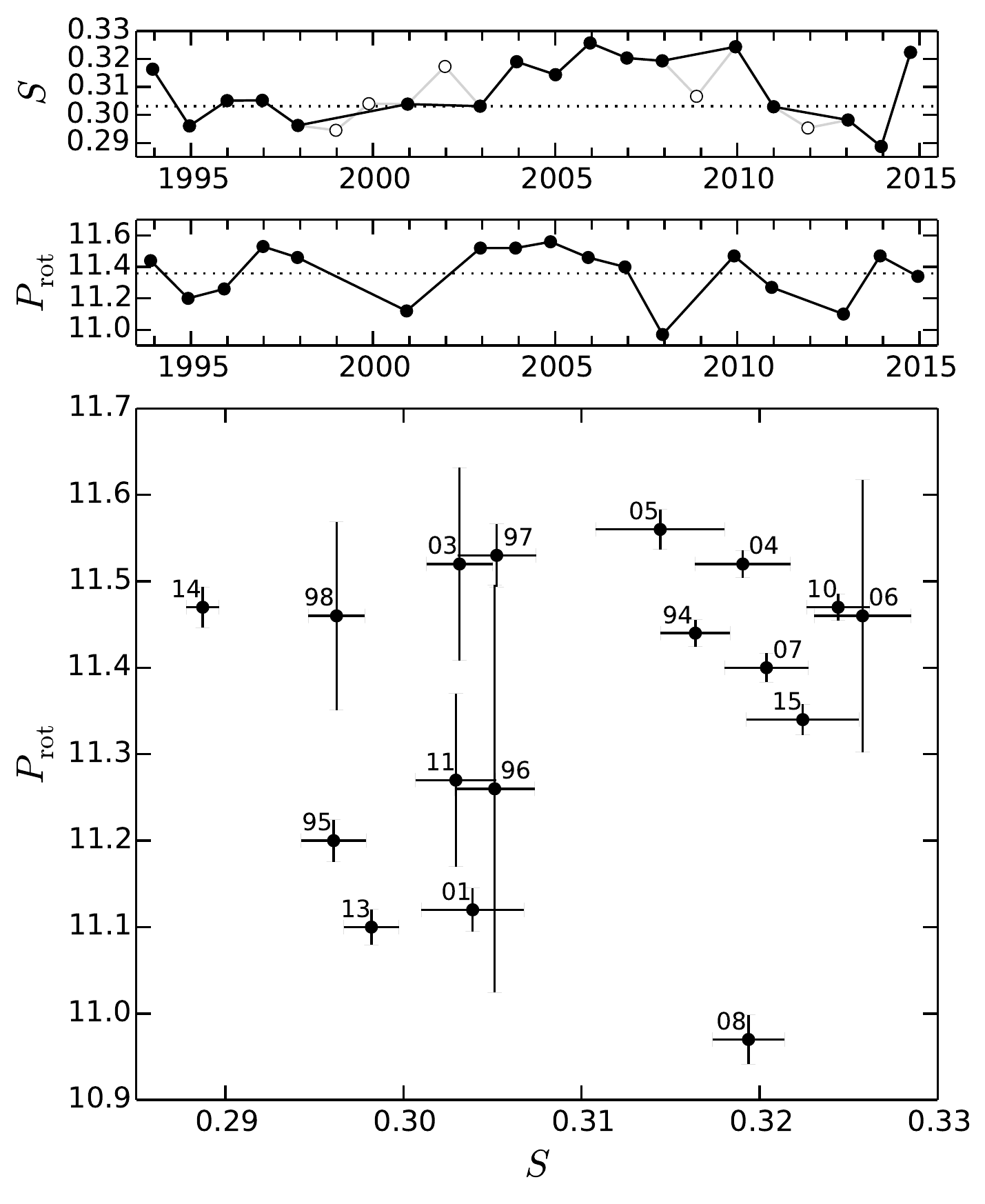}
    \caption{Top two panels: time series of seasonal rotation period
      measurements beneath seasonal mean $S$-index time series, for
      comparison.  Filled (open) circles are measurements for seasons
      in which a rotation period detection was successful
      (unsuccessful).  A horizontal dotted line marks the grand mean
      for the whole time series.  Bottom panel: seasonal
      activity-rotation correlation plot, with error bars representing
      1-$\sigma$ uncertainty in the rotation period and seasonal mean
      $S$ value.  Data points are annotated with the two-digit year.}
    \label{fig:rotation}
\end{figure}

\revone{
The time series of rotation measurements and the seasonal
activity-rotation relationship are shown in Figure \ref{fig:rotation}.
$P_{\rm rot}$ vs. $S$ is a kind of pseudo-butterfly diagram in
the absence of spot latitude information, which would demonstrate
different morphologies under different migration patterns
\citep{Donahue:1993:thesis}.  For example, in the Sun, if rotation
were measured by tracking spots during the cycle we would expect a
long rotation period and low activity at solar minimum (high latitude
spots), transitioning to shorter rotation periods and higher activity
until solar maximum (mid-latitude spots), and finishing with still
shorter rotation periods as activity wanes (near-equator spots).  A
variety of morphologies were observed in \cite{Donahue:1993:thesis},
including the anti-solar case.  From Figure \ref{fig:rotation}, we see
that for HD 30495 in general rotation is slow
when activity is high (with the exception of the 2008
season), but when activity is low both long and short rotation periods
are seen.  Tracing points in chronological order reveals no clear
pattern, with seasons transitioning from quadrant to quadrant in the
diagram.  Near the maxima of cycle 3 ($t = 2005.8$, ref Table
\ref{tab:cycles}) there is a cluster of four seasons with slow
rotation, 2004-2007.  The 1997-1998 rotation was also slow, which
occurred just before the minima at the start of cycle 3 ($t =
1998.4$), however the 2011 and 2013 rotation before the start of cycle 4
($t = 2013.9$) was relatively fast.  The lack of coherent structure
in Figure \ref{fig:rotation} may be an indication that large spots
which make the detection of $P_{\rm rot}$ possible are not restricted
to a narrow range of latitudes as for the Sun.}

Line-of-sight inclination is an important factor for interpreting the
stellar rotation and activity data.  We estimated the inclination
using the $v \sin i$ measurement of \cite{Gaidos:2000}, together with
$v = 2 \pi R / P_\rot$, where $R$ was derived using $g \propto M/R^2$
and the mass and surface gravity estimates of \cite{Baumann:2010} (see
Table \ref{tab:prop}), as well as our rotation measurement.  We
obtained $\sin i = 1.0 \pm 0.2$, indicating an equator-on view of the
star, but the large uncertainty giving a one-sigma range of $i \gtrsim
55.4^\circ$.  This perspective provides a best-case scenario for
measuring rotation with spot transits, as well as a ``solar-like''
view of the activity cycles.

Finally, using the age-rotation relationship given in equation (3) of
\cite{Barnes:2007} along with our mean rotation period, we obtain an
age of $970 \pm 120$ Myr.  This revises the $t_{\rm gyro}$ from Table
3 of that work, which was based on a lower estimate for the rotation
period.

\section{Discussion}
\label{sec:discussion}

We have observed quasi-periodic signals with representative values of
$P_\Plong = 12.2 \pm [3.0]$ yr and $P_\Pshort = 1.67 \pm [0.35]$ yr in the
chromospheric activity of the fast-rotating ($P_\rot = 11.36 \pm 0.17$
days) solar analog HD 30495.  A simple three-component sine cycle
model shows three full cycles in the time series, each with varying
duration and amplitude.  This, combined with the improved
signal-to-noise from our longer time series, has allowed us to  demonstrate the
``pronounced periodicity'' that was previously lacking to classify
this star as cycling in the \cite{Baliunas:1995} study.

Taking the ratio of these periodicities to the rotation period, $n =
P_\cyc / P_\rot$, we find $n_\Plong \approx 400$ rot/cyc and
$n_\Pshort \approx 50$ rot/cyc respectively, which closely correspond
to the ``active'' and ``inactive'' sequences found in the stellar
sample of \cite{BohmVitense:2007}.  What is remarkable from that work
is that the Sun appeared to be a unique outlier in that sample, with
$n_{\Plong,\Sun} \approx 150$, and taking the solar QBO period as 2
years, $n_{\Pshort,\Sun} \approx 30$.  Why does the Sun appear as an
outlier with respect to the stars?  One explanation could be that the
relatively small sample of 21 stars in \cite{BohmVitense:2007} is
insufficient to show the full picture of the relation between $P_\cyc$
and $P_\rot$, and further data will simply erase the observed trends.
\revone{Indeed, the sample of \cite{Saar:1999}, which is a superset of
  the Bohm-Vitense sample, includes a few neighboring points for
  the Sun on log plots of quantities proportional to $P_\cyc$ and
  $P_\rot$.}  Our datum for HD 30495, however, \revone{is decidedly
  not a neighbor of the Sun on these plots, while it agrees well with
  the trend set by stars on the active branch}.  Both the Sun and this
star have a similar time scale for the observed long and short
periodicities, yet rotation,
$\overline{\Omega}_\star/\overline{\Omega}_\Sun \approx 2.3$, is very
different.  This poses a serious problem for Babcock-Leighton flux
transport dynamo models whose time scale is determined by the
meridional flow.  3D hydrodynamic models show that meridional flow
speed decreases with increased rotation rate
\citep{Ballot:2007,Brown:2008}, but kinematic mean-field dynamos with
one meridional flow cell have cycle times proportional to the flow
speed and hence slower cycles with faster rotation.  \cite{Jouve:2010}
investigated this problem, finding that multiple meridional flow cells
in latitude were needed to make cycle period proportional to rotation.
Unfortunately, observations to determine whether this indeed occurs in
fast-rotating stars such as HD 30495 are nearly beyond imagination due
to the extremely slow flow speeds, of order $\sim$10 m/s on the Sun.

{ \revonecolor
We also compared relative amplitudes $A_{\rm cyc} = \Delta S /
\overline{S}$ of the long and short variations to the stellar ensemble
in \cite{Saar:2002}.  We found the \emph{average} cycle amplitude to be
lower than the lowest star in that sample, though not excessively far
from other active-branch stars, while the \emph{maximum} amplitude
(cycle 3) matches closely with the other active branch stars of
similar $B-V$ color.  That the average value is low is likely due to
differences in methods used by \cite{Saar:2002} (and \cite{Soon:1994},
the origin of part of the sample) which estimated peak-to-peak
amplitude using the entire 25-year time series of the MWO sample.  In
addition, our method used the 3-component cycle model and a global
minimum defined by the lowest seasonal median in the time series,
which by design filters out the variance from high-frequency
components, resulting in a lower estimate in amplitude.
}

HD 30495 is nearly rotation and cycle-degenerate with the K2 star
$\epsilon$ Eridani, which has $P_\rot = 11.1$ days and $P_\Plong =
12.7$ yr, but a longer $P_\Pshort = 2.95$ yr \citep{Metcalfe:2013}.
In this comparison, \revone{the experimental variables include the
depth of the convection zone, with stellar structure models
predicting a deeper convective region for the cooler K star, as well
as the average convective velocity.  Apparently, these factors alone
are not enough to prescribe a substantially different long-term
cycle period.}

In most dynamo theories, differential rotation is the driving force of
the $\Omega$-effect, responsible for turning poloidal magnetic flux
into toroidal flux \citep[e.g.][]{Babcock:1961}.  Stars with greater
differential rotation would be expected to ``wind up'' the field
faster, and we should reasonably expect a shorter cycle time, as at
least half of the process would be faster (the other half being the
return of toroidal field to poloidal).  Using our range of rotation
period detections, the total measured surface rotational shear is
$\Delta \Omega = 1.67 \pm 0.15$ degrees/day.  Then, using the the
solar surface differential rotation result of \cite{Snodgrass:1990},
we calculate the equator-to-pole total shear for comparison, finding
$\Delta\Omega_\star / \Delta\Omega_\Sun \geq 0.40 \pm 0.03$, which is a lower
bound due to the unknown latitude ranges causing the rotation signal
on the star.  However, if one is prepared to assume that the spot
latitudes of HD 30495 never form above 45\degree, as for the Sun, then
differential rotation can be compared in terms of the solar shear from
the equator to 45\degree, giving $\Delta\Omega_\star /
\Delta\Omega_{45, \Sun} \geq 1.02 \pm 0.09$.  Both results allow the
possibility that HD 30495 and the Sun have equivalent surface
differential rotation, which might help to explain their similar cycle
characteristics.  Asteroseismic measurements of HD 30495 may be able
to put tighter limits on $\Delta \Omega$ \citep{Gizon:2004,Lund:2014}.

{ \revonecolor
Our time-frequency analysis does not indicate coupling between the
intermittent short-period variability and the long-period cycle for
this star.  This is in contrast to the solar QBO, whose amplitude is
strongly modulated by the 11-yr cycle \citep{Bazilevskaya:2014}.  For
HD30495, the amplitudes of the short-period variations are at times
larger than the amplitude of the long-term cycle, bringing into
question which component of the variability is more fundamental to its
dynamo.  The absence of correlation between the two periodicities may
be an indication that they are of a fundamentally different nature.
It would be interesting to know at what time scale global magnetic
field polarity reverses in this star, if indeed it does reverse.  The
$S$ time series during the densely-sampled SMARTS era shows a
convincing sinusoidal variation, perhaps as convincing as the
long-term trend.  A campaign of Zeeman Doppler Imaging
measurements spaced over at least a four year period should be able to
determine if the short-period variability is polarity-reversing as
well.
}

To the best of our knowledge, this is the first work to separate and
characterize \revone{the amplitudes and durations of} individual
cycles from a stellar activity proxy.  There are rich opportunities in
this direction to explore the \revone{variability \emph{of the cycles
    themselves}, as well as} differences in stellar behavior during
times of minima and maxima, which in turn can provide additional
constraints for dynamo models.  Already, the periodic signals measured
here, together with the global properties collected in Table
\ref{tab:prop}, present a well-characterized object to study with
dynamo models.  \revone{The existence of two significant time scales
  of variability in activity poses an additional modeling constraint.}
This bright object is also a prime candidate for future asteroseismic
observations, which can further constrain its mass, radius, rotation
profile, and depth of the convection zone \citep{Metcalfe:2009}.
Successful modeling of such well-described targets will hopefully lead
to improved understanding of the dynamo process.

\begin{acknowledgements}
We are thankful to Willie Soon \revone{and Sallie Baliunas} for
providing data from the Mount Wilson survey.  Thanks also to Jorge
Melendez, Fabricio Freitas, and Megan Bedell for providing
measurements derived from the HARPS spectrograph (program ID
188.C-0265).  The authors thank Lou Boyd, Director of Fairborn
Observatory, for his tireless efforts in support of our automated
telescopes.  Thanks to Piet Martens and Phil Judge for the useful
discussions.  \revone{We also thank the anonymous reviewer for their
thoughtful comments, which led to substantial improvements in this
manuscript.  This research has made use of the SIMBAD database,
operated at CDS, Strasbourg, France.  This research also made use of
Astropy, a community-developed core Python package for astronomy
(\url{http://astropy.org}).} R.E. is supported by the Newkirk
Fellowship at the High Altitude Observatory.  T.S.M. was partially
supported by the Stellar Astrophysics Centre, which is funded by the
Danish National Research Foundation under grant DNRF106.
G.W.H. acknowledges support from NASA, NSF, Tennessee State
University, and the State of Tennessee through its Centers of
Excellence program.
\end{acknowledgements}

\section*{Appendix: Observations}

\begin{deluxetable*}{cccc}[htp]
\tabletypesize{\small}
\tablecolumns{4}
\tablewidth{0pt}
\tablecaption{S-INDEX OBSERVATIONS OF HD 30495}
\tablehead{
  \colhead{(HJD $-$ 2,400,000)} & \colhead{$S$} & \colhead{$\sigma_S$} & \colhead{Instrument}
}
\startdata
41946.486 & 0.29000 & 0.00897 & MWO \\
41960.475 & 0.31270 & 0.00897 & MWO \\
41961.498 & 0.30230 & 0.00897 & MWO \\
41962.484 & 0.30370 & 0.00897 & MWO \\
42016.505 & 0.32040 & 0.00897 & MWO \\
...       & ...     & ...     & ...
\enddata
\label{tab:obs_S}
\end{deluxetable*}

\begin{deluxetable*}{ccccccc}[htp]
\tabletypesize{\small}
\tablecolumns{7}
\tablewidth{0pt}
\tablecaption{PHOTOMETRIC OBSERVATIONS OF HD 30495}
\tablehead{
  \colhead{Date} & \colhead{(P$-$C1)$_{b}$} & \colhead{(P$-$C1)$_{y}$} & \colhead{(P$-$C2)$_{b}$} & \colhead{(P$-$C2)$_{y}$} & \colhead{(C2$-$C1)$_{b}$} & \colhead{(C2$-$C1)$_{y}$} \\
  \colhead{(HJD $-$ 2,400,000)} & \colhead{(mag)} & \colhead{(mag)} & \colhead{(mag)} & \colhead{(mag)} & \colhead{(mag)} & \colhead{(mag)}
}
\startdata
49239.9956 &  $-$.4011 & $-$.2222 &  $-$.2232 &  $-$.2816 & $-$.1779 &  .0593 \\
49248.9742 &  $-$.3969 & $-$.2197 &  $-$.2234 &  $-$.2805 & $-$.1734 &  .0607 \\
49249.9655 &  $-$.4028 & $-$.2237 &  $-$.2245 &  $-$.2825 & $-$.1783 &  .0588 \\
49250.9688 &  $-$.3988 & $-$.2209 &  $-$.2230 &  $-$.2768 & $-$.1758 &  .0559 \\
49251.9645 &  $-$.3909 & $-$.2144 &  $-$.2166 &  $-$.2708 & $-$.1743 &  .0564 \\
49252.9611 &  $-$.3925 & $-$.2177 &  $-$.2175 &  $-$.2736 & $-$.1749 &  .0558 \\
...        &  ...      & ...      &  ...      &  ...      & ...      & ...
\enddata
\label{tab:obs_by}
\end{deluxetable*}

The observational data used in this study are available in the online
version of this publication.  Table \ref{tab:obs_S} shows a sample of
the nightly $S$ measurements available from MWO, SSS, SMARTS, CPS, and
HARPS shown in Figure \ref{fig:timeseries} (a) and (b).  Note that the
equal-mean calibration described in Section \ref{sec:analysis} is
applied to these data; the data can be returned to its original
calibration by dividing by the constants described there.  Table
\ref{tab:obs_by} gives a sample of the nightly differential photometry
measurements in the Str\"{o}mgren $b$ and $y$ bands.  Letting $c_i$ be
the $i$th column of Table \ref{tab:obs_by}, the $\Delta(b + y)/2$ time
series of HD 30495 in Figure \ref{sec:analysis}(c) is obtained using:

\[
\Delta (b + y)/2 _{\rm HD30495} = \frac{1}{2} \left( \frac{c_2 + c_4}{2} + \frac{c_3 + c_5}{2} \right)
\]

and subtracting the mean.  The difference between the comparison stars
is given by:

\[
\Delta (b + y)/2 _{\rm comps} = \frac{c_6 + c_7}{2}
\]

The seasonal means of this series are shown as white squares in Figure
\ref{fig:timeseries}(c).  Finally, the color difference series $\Delta
(b-y)$ of Figure \ref{fig:timeseries}(d) is given by:

\[
\Delta (b - y)_{\rm HD30495} = \frac{c_2 + c_4}{2} - \frac{c_3 + c_5}{2}
\]

\bibliography{hd30495}
\bibliographystyle{apj}

\end{document}